\documentclass[aip,jcp,reprint,floatfix,showpacs,tighten]{revtex4-1}
\usepackage[dvips]{epsfig}
\usepackage[version=3]{mhchem} 
\newcommand{\dd}{\textrm d}

\newcommand{\vect}[1]{\mathbf{#1}}
\newcommand{\tw}{\text{tw}}
\newcommand{\tens}[1]{\mathsf{#1}}
\newcommand{\un}[1]{\ensuremath{\unskip\,\mathrm{#1}}}

\begin{document}
\title{A two-dimensional nematic phase of magnetic nanorods}\thanks{Originally published in \textit{J. Chem. Phys.} \textbf{140}(10), 104904 (2014)}\homepage[\linebreak]{http://dx.doi.org/10.1063/1.4867790}
\author{Kostyantyn Slyusarenko}
\author{Doru Constantin}
\email{doru.constantin@u-psud.fr}
\author{Patrick Davidson}
\affiliation{\mbox{Laboratoire de Physique des Solides, Universit\'{e} Paris-Sud, CNRS, UMR 8502, 91405 Orsay, France.}}

\begin{abstract}
We report a hybrid mesophase consisting of magnetic nanorods confined between the non-ionic surfactant bilayers of a lamellar phase. The magnetic field-induced ordering of the nanorods was measured experimentally and modeled by a two-dimensional Onsager theory including the third virial coefficient. The nanorods are strongly confined in layers, with no orientational coupling from one layer to the next. At high volume concentration they exhibit spontaneous in-plane orientational ordering and form a stack of independent two-dimensional nematic systems. This isotropic-nematic transition is first-order.
\end{abstract}

\pacs{82.70.Dd, 61.30.-v, 61.05.cf, 78.20.Fm}


\maketitle

\section{Introduction} \label{sec:intro}

Ordered phases in reduced dimensions are fascinating systems, with physical properties that can be qualitatively different from those of their three-dimensional counterparts. One of the most (intuitively) straightforward examples would be the two-dimensional nematic phase, with quadrupolar orientational order but no positional order. This deceivingly simple system has been the topic of extensive theoretical \cite{Chen:1993,vanderSchoot97} and numerical \cite{Lagomarsino:2003,Martinez06,Vink07} work, with a particular emphasis on the first- or second-order character of the isotropic-nematic transition.

There are however surprisingly few experimental realisations of such phases. The first qualitative observation involved rigid phospholipid tubules at the air-water interface \cite{Fisch:1994}. Similar results were obtained for Langmuir-Blodgett films of nanorods \cite{Kim:2001,Li:2003}, but the particles could only be observed after deposition on a substrate. Another strategy relies on inserting DNA molecules within stacks of neutral \cite{Pott02,Pott03} or charged \cite{Bouxsein:2011} lipid bilayers. These disordered mixtures were studied by X-ray scattering and the nematic phase was identified indirectly, by analyzing the shape of the DNA interaction peak. It is not clear whether these are purely two-dimensional phases or whether there is some interlayer coupling. We emphasize that the nematic order parameter was not measured in any of the systems above.

In this paper, we present a system where rigid magnetic nanorods are inserted into a soft lamellar matrix of nonionic surfactant. The resulting phase has three key advantages: 
\begin{itemize}
\item{} It is easily aligned by thermal treatment, so that in the small-angle x-ray scattering (SAXS) images we can clearly discriminate between the direction of the smectic director and that in the plane of the layers. We can thus explore the anisotropy of the phase.
\item{} The nanorods have magnetic properties, hence their orientation can be controlled using an applied magnetic field.
\item{} When the phase is aligned in homeotropic anchoring, we can perform optical birefringence measurements (in particular under magnetic field), which yield a completely independent estimate of the orientational order of the particles.
\end{itemize}
The formulation of the phase \cite{Beneut08} and its structural study \cite{Constantin10} have already been presented. Here, we are concerned with the orientational order of the particles under magnetic field, that we describe using an Onsager-type model (up to the third virial coefficient), accounting for the electrostatic effects and for an applied magnetic field.

We conclusively demonstrate a first-order phase transition between two-dimensional isotropic and nematic phases, settling a long-standing debate in the theoretical and numerical literature \cite{vanEnter:2002,Lagomarsino:2003,Martinez06,Vink07,Almarza:2010}.

\section{Model and analysis} \label{sec:model}

The nanorods can be seen as parallelepipeds, with dimensions $L \times W \times H$ and volume $V = L W H$. They bear a permanent magnetic moment $\mu$ along their long axis and exhibit negative susceptibility anisotropy $\Delta \chi = \chi _{\|} - \chi _{\bot} < 0$. For simplicity, we will describe them in the following as cylinders of length $L$ and diameter $D=\sqrt{4WH/\pi}$. 
 
The particle orientation is quantified by the distribution $f(\Omega)$, such that a fraction $f(\Omega) \dd \Omega$ of particles have their long axes $\vect{e}$ within the solid angle element $\dd \Omega$ around direction $\Omega$.

Our main result concerns the orientational distribution of the particles under confinement, but we start the discussion with their behavior in aqueous solution, a configuration that we will use as reference.

In isotropic solution at a volume concentration $\phi$ and in the presence of a magnetic field (taken along the polar axis, $\vect{B} = B \hat{z}$), $f(\Omega)$ can be described by an extended Onsager model \cite{Vroege92, Lemaire04b}:
\begin{equation}
\begin{split}
f(\Omega) = \frac{1}{A} \exp & \left [ \vphantom{\frac{32 C}{\pi}} KB \cos \theta +JB^2P_2(\cos\theta) \right. \\
& \left. - \frac{32}{\pi} \frac{\phi}{\phi^*} \int \left| \sin \gamma(\Omega, \Omega') \right| f(\Omega') \dd \Omega' \right ]
\end{split}
\label{eq:Onsager}
\end{equation}
with $K = \frac{\mu}{k_BT}$ and $J = \frac{\Delta \chi V}{3 \mu_0 k_B T}$, where $P_2$ is the Legendre polynomial of the second order, $\gamma$ is the angle between two rods oriented along directions $\Omega = (\theta,\varphi)$ and $\Omega ' = (\theta ',\varphi ')$, while
\begin{equation}
\phi^*=\frac{4D^2L}{(D+X)(L+X)^2}
\label{eq:phistar}
\end{equation}
is the spinodal concentration (above which the isotropic state is absolutely unstable), $X$ is an effective length describing the electrostatic interaction between rods \cite{Vroege92} ($X = (\ln A' + \gamma + \ln 2 - 0.5) / \kappa$, with $A'$ the amplitude of the electrostatic interaction, $\gamma \approx 0.577$ Euler's constant, and $\kappa^{-1}$ the Debye screening length), and $A$ is a normalization constant imposing $\int f(\Omega) \dd \Omega = 1$.

Equation \eqref{eq:Onsager} is implicit in $f(\Omega)$ and must be solved numerically. The resulting order parameter $S$ is used to fit the experimental data. 

When the particles are confined in the lamellar phase (with interlayer spacing $d$), we consider that within the same layer they have the same effective interaction as in solution so we model them using the 2D analogue of Equation~\eqref{eq:Onsager}, but this time the polar axis is along the director $\hat{z}\| \vect{n}$ and the magnetic field is applied in the plane of the layers $\vect{B} \bot \vect{n}$ :
\begin{equation}
\begin{split}
&f(\varphi)  = \frac{1}{A} \exp  \left[ \vphantom{\frac{32 C}{\pi}} KB \cos \varphi +JB^2P_2(\cos\varphi) \right. \\
& - \eta \int_0^{2 \pi}  k_2(\varphi, \varphi') f(\varphi') \dd \varphi' \\
& - \eta^2 \int_0^{2 \pi} \int_0^{2 \pi}  k_3(\varphi, \varphi', \varphi'') f(\varphi')
f(\varphi'') \dd \varphi' \dd \varphi'' \\
& +\left. \eta \frac{M}{k_BT} \int_0^{2 \pi}  \cos^2(\varphi - \varphi') f(\varphi') \mathrm d \varphi' \right],
\end{split} 
\label{eq:Onsager2D}
\end{equation}
where the in-plane angle $\varphi$ is measured with respect to the orientation of the magnetic field and $\eta$,
\begin{equation*}
		\eta=\cfrac{4\phi d (L+X) (D+X)}{\pi D^2 L},
\end{equation*}
is the effective surface fraction. By using $D$, we implicitly consider that the particles can rotate freely along their long axis, as in the isotropic solution. If they were to adopt a particular configuration, for instance with the largest face $(L \times W)$ in contact with the surfactant bilayer, the values of $X$ and $\eta$ would be slightly changed. However, we have no experimental evidence for or against this assertion and, for simplicity's sake, we conserve the ``cylindrical'' model used above.

In~\eqref{eq:Onsager2D} the third and fourth terms on the right-hand side describe the second and third virial coefficients of the Onsager interaction. 
The coefficient $k_2(\varphi, \varphi')$ has an analytical form,
\begin{equation*}
		k_2(\varphi, \varphi') = \left( \lambda + \cfrac{1}{\lambda} \right) 
				| \sin(\varphi - \varphi')| 
				+ 2[1+|\cos(\varphi - \varphi')|] ,
\end{equation*}
where $\lambda = \cfrac{L+X}{D+X}$ is the effective aspect ratio, and the coefficient $k_3(\varphi, \varphi', \varphi'')$ is calculated numerically \cite{Martinez06}. Note that Onsager's argument for neglecting higher-order virial terms does not apply in 2D \cite{Straley71, Martinez06}, especially for rods of moderate aspect ratio (such as those used in our experiments). We therefore kept the third term in the virial expansion.

As presented so far, the 2D model considers that particles in the same plane interact by the effective hard-core model measured in solution, while particles in different layers do not interact at all. This latter feature is indispensable for having a true 2D phase, so it should be thoroughly checked. That is why we added in~\eqref{eq:Onsager2D} a possible orientational coupling\footnote{In principle, one could also have a positional coupling between layers. We do not include this type of term, because here we concentrate on the orientational order and because an extensive X-ray scattering investigation showed no characteristic sign (peak with $q_z \neq 0$ and $q_r \neq 0$) of such coupling.} between the nanorods in neighboring layers, described by the last term (with $M$ the coupling amplitude). Below, we discuss possible physical origins for such a coupling, estimate its strength and show that it is negligible.

\subsection{X-ray scattering}\label{sec:SAXS}

\begin{figure}[htbp]
\includegraphics[width=0.45\textwidth,angle=0]{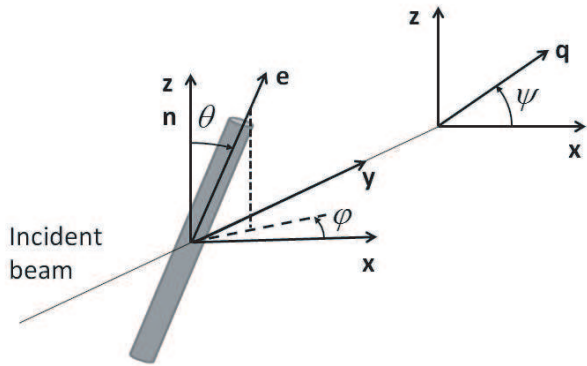}
\caption{Experimental geometry. The scattering object is a rod with orientation $\vect{e}$ given by the polar angle $\theta$ (measured with respect to the nematic director $\vect{n} \| \hat{z}$) and the azimuthal angle $\varphi$. The incident beam is along the $\hat{y}$ axis. The detector records in the $(x,z)$ plane (normal to the incident beam) the scattered intensity $I(q,\psi)$ given by Equation \eqref{eq:Iqpsi}}\label{fig:geom}
\end{figure}

The X-ray scattering form factor of an individual particle is:

\begin{equation}
F(q, \beta, L, D) =F_0 \, \text{sinc} \left( \frac{qL\cos\beta}{2} \right)
\left( \frac{J_1(qD/2\sin\beta)}{qD/2\sin\beta} \right)
\end{equation}
where $\beta$ is the angle between the long axis of the rod $\vect{e}$ and the scattering vector $\vect{q}$: $\cos \beta = \cos \theta \sin \psi + \sin \theta \cos \psi \cos \varphi$ (see also Figure \ref{fig:geom}).

For a monodisperse population, the scattering signal is given by \cite{Lemaire02c}:
\begin{equation}
I(q,\psi) = 2 \int_0^{2\pi} \dd \varphi \int_0^{\pi /2}
f_{MS}(\theta)F^2(q,\beta) \sin (\theta ) \dd \theta
\label{eq:Iqpsi}
\end{equation}
while in the general (polydisperse) case Equation \eqref{eq:Iqpsi} must be averaged over a radius distribution $g(D)$ (the polydispersity in the length $L$ does not contribute over the available $q$-range.)
 In \eqref{eq:Iqpsi} we approximate the orientational distribution of the rods by the analytical Maier-Saupe form:
\begin{equation}
f_{MS}(\Omega) = \frac{\exp(m\cos^2 \theta)}{Z}
\label{eq:MS}
\end{equation}
where $Z$ is a normalization constant and fitted the data with only one parameter $m$. The order parameter $S$ can be expressed analytically as a function of $m$.

\subsection{Optical birefringence}\label{sec:Opt}

The birefringence of the samples can generally be expressed as:
\begin{equation}
\Delta n = \Delta n_{\text{sat}} S(B) \phi
\label{eq:Dn}
\end{equation}
where the field-dependent order parameter $S(B)$ is the uniaxial one $S$ for the case of water suspensions and the in-plane component $\frac{2}{3}P$ of the biaxial order when the particles are confined within the lamellar matrix. 
$\Delta n_{\text{sat}}$ is the specific birefringence, depending on the geometry of the particles and on their dielectric permittivity tensor.

\section{Results and Discussion} \label{sec:res}

From electron microscopy and X-ray diffraction measurements, the average particle dimensions are:
\begin{equation}
L = 315 \un{nm} \times W = 38 \un{nm} \times H = 18 \un{nm} \, ,
\label{eq:size}
\end{equation}
and the relative polydispersity is of the order of 0.3 \cite{Constantin10}. They correspond to cylinders with a diameter $D = \sqrt {4WH /\pi} = 29.5 \un{nm}$.

\subsection{In solution}\label{sec:sol}

Under a magnetic field $\vect{B}$, aqueous solutions of nanoparticles become uniaxial, with a director $\vect{n} \| \vect{B}$ and an order parameter $S$ that can be positive or negative, depending on the field amplitude \cite{Lemaire04b}.

\begin{figure}[htbp]
\includegraphics[width=0.5\textwidth,angle=0]{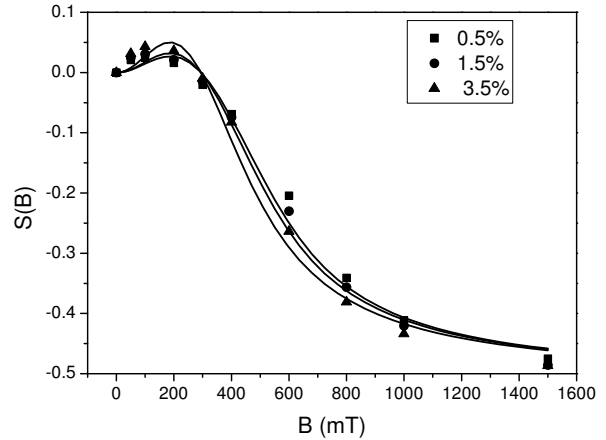}
\caption{Order parameter $S(B)$ (determined from the SAXS images) for the aqueous suspensions of nanorods at various concentrations $\phi$ (symbols). The curves are fits derived from Equation~\eqref{eq:Onsager} with parameter values~\eqref{eq:const}}\label{fig:SBaq}
\end{figure}

From the SAXS images we determined the order parameter $S(B)$ as a function of the field (Figure~\ref{fig:SBaq}) by fitting the data to Equation~\eqref{eq:Iqpsi} and using for the orientation distribution the Maier-Saupe form \eqref{eq:MS}. The polydispersity $g(D)$ is estimated from the scattering spectrum of a dilute suspension in the absence of the field.

The order parameter is then computed over the distribution $f (\theta)$ determined from the complete model~\eqref{eq:Onsager} as a function of $B$ and $\phi$, yielding estimates for the material constants $K$ and $J$.  

The spinodal concentration $\phi^* = 7.1 \%$ was estimated  from the independent measurement of the phase diagram of the aqueous suspension in zero field: $\phi^* \approx 0.95 \phi_{N}^{3D}$, where $\phi_{N}^{3D}$ is the volume concentration of the nematic phase at coexistence (in aqueous solution). From \eqref{eq:phistar} we then obtain $X=72 \un{nm}$.

\begin{figure}[htbp]
\includegraphics[width=0.5\textwidth,angle=0]{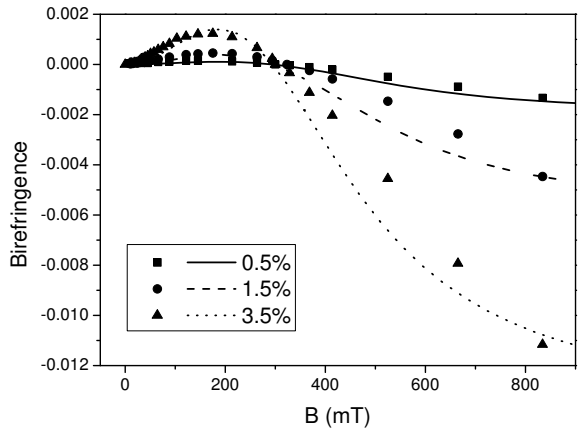}
\caption{Birefringence $\Delta n (B)$ of the aqueous suspensions of nanorods at various concentrations $\phi$ (symbols). The curves are fits derived from Equations~\eqref{eq:Dn} and \eqref{eq:Onsager} with parameter values~\eqref{eq:const}}
\label{fig:DnBaq}
\end{figure}

We also measured the birefringence $\Delta n (B)$ of the solutions (Figure \ref{fig:DnBaq}) and fitted it to Equation~\eqref{eq:Dn}, which involves the additional constant $\Delta n_{\text{sat}}$.

Finally, we obtain the material constants:

\begin{subequations}
\label{eq:const}
\begin{align}
K &= 8.2 \pm 0.5 \un{T^{-1}} \label{eq:constK}\\
J &= -13.8 \pm 1 \un{T^{-2}}  \label{eq:constJ}\\
\Delta n_{\text{sat}} &= 0.8        \label{eq:constnsat}\\
\phi^* &= 7.1 \% \, \Rightarrow \, X = 72 \un{nm} \label{eq:constC}
\end{align}
\end{subequations}
that we use alongside the particle dimensions~\eqref{eq:size} to study the hybrid system (particles inserted in the lamellar matrix).

\subsection{In the lamellar phase}\label{sec:La}

The SAXS measurements can be performed in either homeotropic or planar anchoring, giving access to both $S$ and $P$, but the optical birefringence can only be determined in the homeotropic configuration (sensitive to $P$).

In the planar configuration, with $\vect{B} = 0$, for $\phi = 0.5 \%$, $1.5 \%$, and $3.5 \%$ we obtained $S = -0.44$, $-0.46$, and $-0.47$, respectively, very close to the perfect confinement limit $S = -0.5$. We conclude that, even in the absence of the field, the particles are practically contained within the plane of the layers (and the optical axis is along $\vect{n}$). If we now apply a field $\vect{B} \bot \vect{n}$ the system becomes biaxial, with a second order parameter $P$. 

In the following we treat the system as purely two-dimensional, with $S = -0.5$. The particle population is then completely described by the orientation distribution $f (\varphi)$, determined from the model~\eqref{eq:Onsager2D} as a function of $B$, $M$, and $\phi$. The material constants are those measured in aqueous solution~\eqref{eq:const}, the periodicity of the lamellar phase $d= 45 \un{nm}$, \cite{Constantin10} and $M$ is the sole fitting parameter. 

\subsubsection{Isotropic two-dimensional phase}\label{sec:Lalow}
\begin{figure}[htbp]
\includegraphics[width=0.5\textwidth,angle=0]{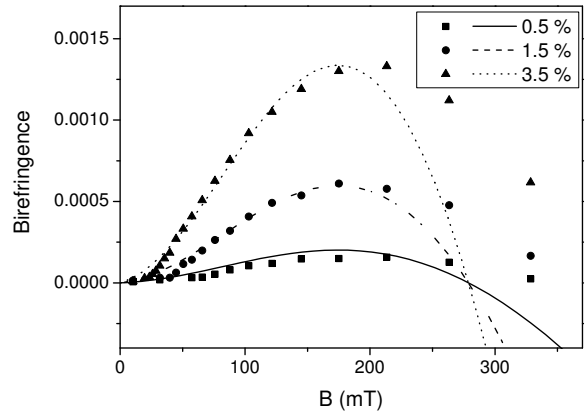}
\caption{Birefringence $\Delta n (B)$ of the suspensions of nanorods in lamellar phase at various concentrations $\phi$ (symbols). The curves are fits derived from Equations~\eqref{eq:Dn} (with $S(B)$ replaced by $(2/3) P(B)$) and \eqref{eq:Onsager2D} with the material parameters in~\eqref{eq:const}, $d= 45 \un{nm}$, and $M=-3 \, k_BT$.}
\label{fig:DnBla}
\end{figure}

\begin{figure}[htbp]
\includegraphics[width=0.5\textwidth,angle=0]{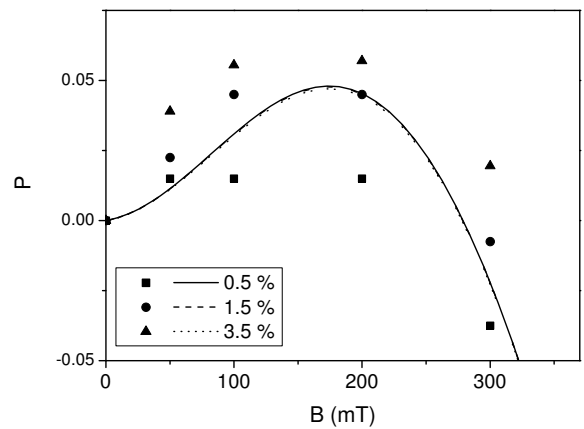}
\caption{Second order parameter $P(B)$ (determined from the SAXS images) of the suspensions of nanorods in lamellar phase at various concentrations $\phi$ (symbols). The curves are fits derived from Equation~\eqref{eq:Onsager2D} with the material parameters in~\eqref{eq:const}, $d= 45 \un{nm}$, and $M=-3 \, k_BT$.}
\label{fig:SBla}
\end{figure}

We describe the birefringence $\Delta n (B)$ using Equation~\eqref{eq:Dn} (with $S(B)$ replaced by $(2/3) P(B)$) coupled with \eqref{eq:Onsager2D}. The simultaneous fit of the three curves yields $M=(-3 \pm 3) \, k_B T$ see Figure~\ref{fig:DnBla}, and the resulting order parameter is compared to that measured by SAXS in Figure~\ref{fig:SBla}. The $M$ values for the individual fits at 1.5 and 3.5~\% are shown in Figure~\ref{fig:Mf}. At 0.5~\% the uncertainty on $M$ is very large, so we excluded this point from the analysis.

\subsubsection{Nematic two-dimensional phase}\label{sec:Lahigh}

The most concentrated system, with $\phi = 8 \, \text{vol \%}$ in the lamellar phase is a biaxial nematic, with a spontaneous second order parameter $P$. It can thus be seen as a stack of two-dimensional nematic layers. To measure $P$ we applied a small magnetic field of $30~\un{mT}$ in the plane of the layers to orient all nematic domains along the field, and measured the dependence of the x-ray scattering intensity $I$ on $\psi$. Fitting $I(\psi)$ by the Leadbetter method~\cite{Leadbetter79}, which takes into account the interaction between nanorods (and is more appropriate at high concentration than the model \eqref{eq:Iqpsi}), we obtained $P = 0.51$.

\begin{figure}[htbp]
\includegraphics[width=0.4\textwidth,angle=0]{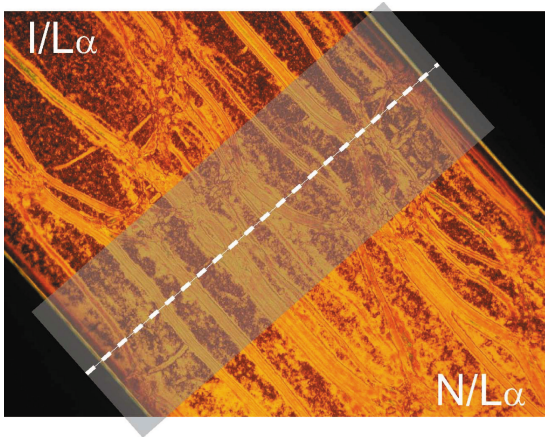}
\caption{Sample at coexistence ($\phi= 6.6\, \text{vol \%}$, $\eta = 0.54$). The dashed line (with a greyed uncertain area) delimits the 2D isotropic phase $I/L_{\alpha}$ (upper left) from the 2D nematic phase $N/L_{\alpha}$ (lower right). The width of the capillary is 1~mm.}
\label{fig:coex}
\end{figure}

Both in aqueous solution and in the lamellar phase the isotropic-nematic transition is first-order, as the two phases can coexist over a certain concentration range. By preparing several samples within this domain and estimating the fraction occupied by each phase we determine the nematic concentration at coexistence $\phi_{N}^{3D}$ (in the aqueous solution) and  $\phi_{N}$ (for the two-dimensional nematic). In our system, these two distinct parameters happen to have the same numerical value, $7.5 \pm 0.5 \, \text{vol \%}$. In the lamellar phase, the isotropic-nematic coexistence range is approximately $\phi= 4.5-7.5 \, \text{vol \%}$, corresponding to $\eta = 0.38-0.62$ \cite{Constantin10}. A sample at coexistence, with $\phi= 6.6\, \text{vol \%}$, i.e. $\eta = 0.54$, is shown in Figure~\ref{fig:coex}. The interface is not sharply defined, mainly due to the presence of the lamellar defects which create arbitrarily shaped domains of the two phases.

The two-dimensional model~\eqref{eq:Onsager2D} does indeed predict a first-order isotropic-nematic transition, with a $\phi_{N}$ that corresponds to the experimental one ($\phi^{\text{exp}}_{N} = 7.5 \pm 0.5 \, \text{vol \%}$) for $M=(0.5 \pm 0.5) \, k_BT$. 
For our most concentrated sample, with $\phi = 8 \, \text{vol \%}$, the same model reproduces the order parameter $P = 0.51$ for $M = (0.07 \pm 0.01) \, k_BT$. These two values correspond to the high-concentration points in Figure~\ref{fig:Mf}.

\subsubsection{Coupling}

\begin{figure}[htbp]
\includegraphics[width=0.5\textwidth,angle=0]{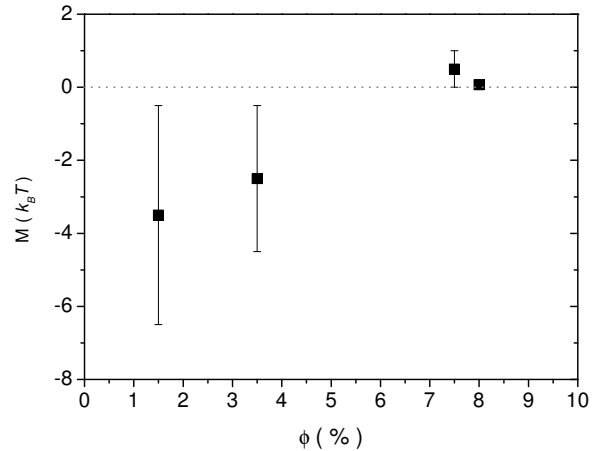}
\caption{Best estimate of $M$ for various concentrations $\phi$. For $\phi = 1.5$ and 3.5 \% we used the birefringence data (Figure \ref{fig:DnBla}). The two high-concentration points are obtained from the transition concentration and from $P(\phi=0.08)$ (see text).}
\label{fig:Mf}
\end{figure}

The coupling coefficient $M$ is not significantly different from zero: at low concentration (in the two-dimensional isotropic phase) $M=(-3 \pm 3) \, k_B T$ from the birefringence data and at high concentration (in the two-dimensional nematic phase) $M=(0.5 \pm 0.5) \, k_BT$ from the transition concentration and $M = (0.07 \pm 0.01) \, k_BT$ from $P(\phi=0.08)$. We conclude that, within the experimental precision, the orientation of particles in different layers is uncoupled, see Figure~\ref{fig:Mf}. Is this result in agreement with the theoretical estimations?

The main difference between the particles in lamellar phase and those in isotropic solution is that the former are constrained to lie in parallel planes, and thus can more easily become parallel. Magnetic and electrostatic interaction might then be stronger than in three dimensions.  We estimate their amplitude in the two-dimensional nematic phase, at $\phi = 8\,\text{vol \%}$:

The magnetic interaction between the permanent dipole moments of the particles $W_\text{dd}$ favors an antiparallel orientation and its magnitude is:
\begin{equation*}
\frac{W_\text{dd}}{k_BT} \approx \frac{\mu_0 \mu^2}{4\pi r^3 k_BT} \approx \frac{\mu_0 K^2 k_BT\phi}{4\pi LDH} = 6 \cdot 10^{-6},
\end{equation*} 
where $r$ is an average interparticle distance. 

The electrostatic twist energy\footnote{This is simply the repulsion between identical and uniformly charged rods placed on top of each other in parallel planes.} per molecule $W_\tw$ (which favors a perpendicular orientation) is given by~\cite{Stroobants86}:
\begin{equation}
\frac{W_\tw}{k_BT}= \frac{\eta h_\tw}{2}
\int\int k_\tw(\varphi,\varphi') f(\varphi) f(\varphi') d\varphi d\varphi', 
\end{equation}
where $h_\tw = (\kappa X)^{-1}$. The coefficient $k_\tw(\varphi,\varphi')$ is:
\begin{equation*}
\begin{split}
& k_\tw(\varphi,\varphi') =  \left( \lambda + \frac{1}{\lambda} \right)
[-|\sin(\varphi-\varphi')|\ln|\sin(\varphi-\varphi')| \\
& -\alpha|\sin(\varphi-\varphi')|], 
\end{split}
\end{equation*}
with 
\begin{equation*}
\alpha = -\frac{\int\int |\sin(\varphi-\varphi')|\ln|\sin(\varphi-\varphi')| d\varphi d\varphi'} 
{\int\int |\sin(\varphi-\varphi')| d\varphi d\varphi'}
\approx 0.3065.
\end{equation*}
The main uncertainty is in evaluating the Debye length: $\kappa^{-1} = 2 - 14 \un{nm}$, where the lower value was measured in a different goethite batch \cite{Lemaire04b} and the higher one corresponds to an ionic strength of 1~mM (the minimum possible at $pH=3$). We then have $W_\tw \approx (4 - 25) \cdot 10^{-3} k_B T$. Both effects are therefore negligible compared to the thermal energy, in agreement with our experimental findings. 

\section{Conclusion} \label{sec:conc}

We formulated stable anisotropic materials where goethite nanorods are confined between the bilayers of a soft lamellar phase. The nanorods form a nematic system with director along the director of the lamellar phase and a negative order parameter $ S{\leq} -0.45$ (to be compared with $S=-0.5$ for perfect confinement). 

At low volume concentration $\phi < 7.5 \, \text{vol \%}$, the system is uniaxial in the absence of a magnetic field and forms a two-dimensional isotropic phase. Under a small magnetic field ($0 < B < 350 \, \text{mT}$) parallel to the layers, the distribution of the nanorods acquires a biaxial character (the long axis is preferentially oriented along the field), with a low order parameter $P$ ($P \leq 0.05$ for $\phi = 3.5 \, \text{vol \%}$). 

At higher concentration $\phi \geq 7.5 \text{vol \%}$ the biaxial order is spontaneous, yielding a two-dimensional nematic phase in the plane perpendicular to $\vect{n}$, with an order parameter $P \simeq 0.51$.

We modelled the magnetic field-induced ordering of the nanorods $P(B)$ by a two-dimensional Onsager theory and obtained good agreement with the experiment. In the simulation we took into account the magnetic properties of the nanorods and the effect of the electrostatic interaction and expanded the free energy to the third virial coefficient. We also included a possible orientational coupling $M$ between nanorods in neighboring layers.

We conclude that $M$ is not significantly different from zero (and, at any rate, much smaller than $k_B T$) and that the phase is a stack of isolated layers. The particles can be modelled as rigid rectangles with aspect ratio 3.8; at low concentration they exhibit a two-dimensional isotropic phase which (on increasing the concentration) undergoes a first-order phase transition to a two-dimensional nematic phase with an order parameter $P \sim 0.5$. This result is in very good agreement with numerical simulations \cite{Cuesta:1990,Martinez06}.

In our analysis, we chose a particular effective shape (rigid rectangles) for the particles, mainly due to its tractability (very important for including the third virial coefficient). Other choices can lead to a different phase diagram, in particular to a second-order phase transition \cite{Martinez05}, in contrast with the experimental findings. We also ignored the effect of polydispersity, which could have profound consequences \cite{Martinez11}. Hopefully, our experimental results will motivate further theoretical and numerical research in the field of two-dimensional ordered phases. 

\begin{acknowledgments}
We acknowledge support from the Triangle de la Physique (project 2011-083T). The ESRF is acknowledged for the provision of beamtime (experiment SC-2393, ID02 beamline). The authors thank C. Chan\'{e}ac for the goethite suspension, P. Boesecke, M. Imp\'{e}ror, A. Poulos, and B. Pansu for assistance with the SAXS experiments, J. Andrieu for performing some of the optical birefringence measurements and L. Navailles for stimulating discussions.
\end{acknowledgments}

\bibliography{Confinement}

\appendix
\section*{Order parameter}
\label{sec:ordparam}
The order parameter tensor $\tens{Q}$ of the phase describes the distribution of the particle direction $\vect{e}$: $Q_{\alpha \beta} = \left \langle \frac{1}{2} (3 \vect{e}_{\alpha} \vect{e}_{\beta} - \delta_{\alpha \beta}) \right \rangle $ (where the average $\left \langle \cdot \right \rangle$ is taken over the distribution $f(\Omega)$) and can be written in the principal axis frame as (\citet[Eq. 7]{Palffy91}):
\begin{equation}
Q_{\alpha \beta} = \left ( \begin{array}{ccc}
-\frac{1}{2}(S-P) & 0 & 0 \\
0 & -\frac{1}{2}(S+P) & 0 \\
0 & 0 & S 
\end{array} \right )
\label{eq:Qab}
\end{equation}

In spherical coordinates, with the polar axis along $\hat{z}$, the values of $P$ and $S$ are related to the distribution function $f(\theta, \varphi)$ via: 
\begin{equation}
\begin{split}
& S = \frac{3}{2} \iint f(\theta, \varphi) \sin \theta \cos^2 \theta \, \dd \theta  \dd \varphi
- \frac{1}{2} \\
& P = \frac{3}{2} 
\iint f(\theta, \varphi) \sin^3 \theta \cos (2 \varphi) \,  \dd \theta \dd \varphi \\
\end{split}
\label{eq:SP}
\end{equation}
For a uniaxial system the distribution $f$ depends only on the polar angle, $f(\Omega) = f(\theta)$, $P=0$ and $S$ is defined via the simplified formula: $S = \frac{3}{2} \int  f(\theta) \sin \theta \cos^2 \theta \, \dd \theta - \frac{1}{2}$. In the ``complete confinement'' case (relevant for the lamellar system), $f(\Omega) = \delta(\theta-\pi/2)f(\varphi)$, $S= -\frac{1}{2}$, and $P$ reduces to: $P = \frac{3}{2} \int  f(\varphi)  \cos (2 \varphi) \, \dd \varphi$.

\end{document}